\documentclass[aps,prx,twocolumn,nofootinbib,10pt]{revtex4-2}
\usepackage{libertine}
\usepackage{times}
\usepackage{comment}
\usepackage{graphicx}
\usepackage{feynmf}
\usepackage{tabularx}
\usepackage{amsmath}
\usepackage{amstext}
\usepackage{amssymb}
\usepackage{xfrac}
\usepackage[colorlinks,citecolor=blue]{hyperref}
\usepackage{graphicx}
\usepackage{amsmath}
\usepackage{amstext}
\usepackage{amssymb}
\usepackage{amsfonts}
\usepackage{longtable,booktabs}
\usepackage{hyperref}
\usepackage{url}
\usepackage{subfigure}
\usepackage{dsfont}
\usepackage{booktabs}
\usepackage{amsbsy}
\usepackage{dcolumn}
\usepackage{amsthm}
\usepackage{bm}
\usepackage{esint}
\usepackage{multirow}
\usepackage{hyperref}
\usepackage{cleveref}
\usepackage{mathrsfs}
\usepackage{amsfonts}
\usepackage{amsbsy}
\usepackage{dcolumn}
\usepackage{bm}
\usepackage{multirow}
\usepackage{color}
\usepackage{extarrows}
\usepackage{datetime}
\usepackage[super]{nth}
\hypersetup{
	colorlinks=magenta,
	linkcolor=blue,
	filecolor=magenta,
	urlcolor=magenta,
}

\newtheorem{theorem}{Theorem}

\newcommand{\Tr}{\text{Tr}}

\begin{document}
\title{Yang-Lee Zeros in Quantum Phase Transition: An Entanglement Perspective}

\author{Hongchao Li}
\affiliation{Department of Physics, University of Tokyo, 7-3-1 Hongo, Tokyo 113-0033, Japan}
\email{lhc@cat.phys.s.u-tokyo.ac.jp}

%\author{Shinsei Ryu}
%\affiliation{Department of Physics, Princeton University, Princeton, New Jersey, 08544, USA}

%\email{shinseir@princeton.edu}

\date{\today}
\begin{abstract}
%We show that Yang-Lee zeros at absolute zero in quantum many-body systems can induce the entanglement transition of the ground states. At the edges of the distribution of Yang-Lee zeros, the entanglement entropy of ground states discontinuously jumps from the subarea law in a gapless phase to the area law in a gapped phase. We demonstrate the general mechanism behind it from the non-interacting fermionic models. To exemplify it, we calculate the distribution of the Yang-Lee zeros for the non-Hermitian Su-Schrieffer-Heeger model at absolute zero. To show the generality of the mechanism, we introduce the simplified Bethe ansatz equation around the ferromagnetic phase transition point and prove a line distribution of Yang-Lee zeros for an arbitrary size. 
%Our study is believed to unveil a universal understanding of the origin of singularity in entanglement entropy at the entanglement transition points.
We study the Yang-Lee theory in quantum phase transitions from the perspective of quantum entanglement in one-dimensional many-body systems. We primarily focus on the distribution of Yang-Lee zeros and its associated Yang-Lee edge singularity of two prototypical models: the Su-Schrieffer-Heeger model and the \emph{XXZ} spin chain. By taking the zero-temperature limit, we show how the Yang-Lee zeros approach the quantum phase transition points on the complex plane of parameters. To characterize the edge singularity induced by Yang-Lee zeros in quantum phase transition, we introduce the entanglement entropy of the ground state to show that the edges of Yang-Lee zeros lead to the ground-state entanglement transition. We further show that our results are also applicable to the general non-interacting parity-time-symmetric Hamiltonians. 
%Our study is believed to unveil a universal understanding of the relation between the Yang-Lee zeros at absolute zero and ground-state entanglement transition.
\end{abstract}
\maketitle
\section{Introduction}
Yang-Lee zeros \cite{PhysRev.87.404,PhysRev.87.410}, defined as the zero points of the partition function of the canonical ensemble, give a mathematical explanation of the nonanalyticity of thermodynamic observables at phase transition points. In the classical ferromagnetic Ising model, Yang and Lee investigated the zeros of the partition function in the presence of an imaginary magnetic field in order to understand the origin of the singularity in the ferromagnetic phase transition when one increases the temperature. The thermal phase transition takes place when Yang-Lee zeros touch the positive real axis on the complex plane of fugacity in the thermodynamic limit. In the vicinity of the edge of the distribution of Yang-Lee zeros, critical phenomena collectively known as Yang-Lee edge singularity emerge, which are accompanied by anomalous scaling laws~\cite{Fisher:1978vn,Kurtze:1979wb,Fisher1980,Cardy:1985ub,Cardy:1989uo,Zamolodchikov:1991tl,BENA2005,Fisher1965,Xu2024}. Yang-Lee theory is applicable to a variety of thermal phase transitions in classical~\cite{Biskup2000,Arndt2000,Blythe2002,Lee2013} and quantum~\cite{Gehlen_1991,Sumaryada:2007uu,Abraham1996,Matsumoto2020,Vecsei2022,Gnatenko2017,Peotta:2021ua,Shen2023,IBena2005,Kist2021,Hongchao2023,Timonin2021,Gennady2022} systems.

Despite the great success of the application of Yang-Lee zeros to explain the origin of the singularity in thermal phase transitions, there exist exceptions in quantum phase transitions such as the topological quantum phase transition~\cite{SSH1980,Ryu_2010}, which are characterized by the change of the topological invariants instead of singularity in thermodynamic quantities.
With the introduction of quantum entanglement, it was discovered that the singularity in a quantum phase transition can be more universally characterized by the singular behavior of the entanglement pattern in ground-state wavefunctions~\cite{Levin2006,Preskill2006,Fidkowski2010,Jian2021}. Therefore, it is expected to universally characterize the Yang-Lee edge singularity in terms of the singularity in quantum entanglement instead of thermodynamic observables in the traditional Yang-Lee theory.
%Entanglement, as one of the most striking phenomena in quantum physics, plays an essential role in understanding the correlation properties between different partitions in quantum systems~\cite{Bell1964}. Given a ground state $|\psi\rangle$ and a bipartition of the space into $A$ and $\bar{A}$ of a quantum system, we can define the reduced density matrix of $A$ as $\rho_A=\text{Tr}_{\bar{A}}[|\psi\rangle\langle\psi|]$ and the entanglement entropy between $A$ and $\bar{A}$ as $S=-\text{Tr}_A[\rho_A\log\rho_A]$ to quantitatively describe the entanglement of the ground state. In quantum phase transitions in one-dimensional cases from the gapless phase to the gapped phase, the entanglement entropy changes from a subarea law to an area law~\cite{Calabrese_2004,Hastings_2007,Grover2011}.
%When a one-dimensional system is endowed with conformal symmetry, the ground-state entanglement entropy of one-dimensional infinitely long chain can be described by subarea law~\cite{Calabrese_2004,calabrese2005entanglement,Calabrese_2009}: $S\sim c/3\log(l/a)$ where $c$ is the central charge and $l$ is the subsystem size. In the gapped one-dimensional systems, one can prove that the entanglement entropy is always an area law: $S=\text{const}$, which does not increase with subsystem size~\cite{Calabrese_2004,Hastings_2007,Grover2011}. The changes of the entanglement behavior is hence called entanglement transitions of ground states.

%Thus, the study of Yang-Lee theory of entanglement transition can uncover hitherto universality in understanding entanglement.

In this work we develop a Yang-Lee theory for quantum phase transitions and focus on how the Yang-Lee zeros approach the phase transition points by taking the zero-temperature limit. First, we investigate the distribution of Yang-Lee zeros in the parameter space of a noninteracting model: the Su-Schrieffer-Heeger (SSH) model~\cite{Chang2019} by generalizing the Rice-Mele coupling~\cite{Rice-Mele} to an imaginary one. We explore the nonunitary criticality of the non-Hermitian SSH model originating from Yang-Lee edge singularity. These quantum critical phenomena are induced by the exceptional points where a nonanalytic excitation spectrum exists. Furthermore, we figure out that the edges of Yang-Lee zeros at absolute zero correspond to the entanglement transition points of the ground states. We also generalize our discussion to general parity-time (PT)-symmetric free-fermion systems to unveil universally unnoticed universality in Yang-Lee edge singularity in terms of transitions in the entanglement entropy. This result indicates that entanglement entropy may serve as a good characterization for the universality class of Yang-Lee edge singularity in quantum phase transitions. In contrast to Ref.~\cite{Jian2021}, we have rigorously  proved the correspondence between the edges of Yang-Lee zeros and ground-state entanglement transitions for all PT-symmetric free-fermion systems rather than entanglement transitions of the long-time steady states of a class of specific models.

We also discuss the Yang-Lee theory of an interacting model: the \emph{XXZ} spin chain. On the complex plane of the anisotropy, we investigate the distribution of Yang-Lee zeros near the ferromagnetic phase transition point. By analyzing the Bethe ansatz of the \emph{XXZ} model, we prove that the Yang-Lee zeros approach the phase transition point in a line for arbitrary system size. The edge of the distribution of Yang-Lee zeros turns out to be the ferromagnetic phase transition point. We also show that the edge of the Yang-Lee zeros corresponds to an entanglement transition. This exact result paves the way for figuring out Yang-Lee zeros in quantum phase transitions of interacting models. Finally, we provide a general argument for the correspondence between Yang-Lee edge singularity and entanglement transition.

\section{Non-interacting Model}We begin from a non-interacting model: the non-Hermitian SSH model. The Hamiltonian in the momentum space is defined as~\cite{Lieu2018,Chang2019}
\begin{equation}\label{eq:second-quantization}
  H = \sum_{k}(c_{kA}^{\dagger}\ \ c_{kB}^{\dagger})H_k\left(\begin{array}{cc}
    c_{kA}\\
    c_{kB}
  \end{array}\right),
\end{equation}
with the Bloch Hamiltonian
\begin{equation}\label{eq:H_SSH}
  H_k =
  \left(\begin{array}{cc}
    i u & w e^{- i k} + v\\
    w e^{i k} + v & - i u
  \end{array}\right),
\end{equation}
where $k$ is the momentum and $c_{kA}^{(\dagger)}$ and $c_{kB}^{(\dagger)}$ represent the annihilation (creation) operators of the fermions with momentum $k$ on the sublattice $A$ and $B$ respectively. The energy spectrum of the Bloch Hamiltonian is $E_k^{\pm}=\pm\sqrt{|v+we^{ik}|^2-u^2}=:\pm E_k$. 
% The right eigenvectors of the Hamiltonian are \cite{Chang2019}
% \begin{equation}
%     |R_{k+}\rangle=\left(\begin{array}{cc}
%     \frac{v_k}{|v_k|}\cos\frac{\phi_k}{2}\\
%     \sin\frac{\phi_k}{2}
%   \end{array}\right),|R_{k-}\rangle=\left(\begin{array}{cc}
%     -\frac{v_k}{|v_k|}\sin\frac{\phi_k}{2}\\
%     \cos\frac{\phi_k}{2}
%   \end{array}\right),
% \end{equation}
% where we define $v_k=v+we^{-ik}$ and $\tan\phi_k=v_k/(iu)$. Those right eigenvectors satisfy $H_k|R_{k\pm}\rangle=E_k^{\pm}|R_{k\pm}\rangle$. Similarly, the left eigenvectors are given by \cite{Chang2019}
% \begin{equation}
%     |L_{k+}\rangle=\left(\begin{array}{cc}
%     \frac{v_k}{|v_k|}\cos^*\frac{\phi_k}{2}\\
%     \sin^*\frac{\phi_k}{2}
%   \end{array}\right),|L_{k-}\rangle=\left(\begin{array}{cc}
%     -\frac{v_k}{|v_k|}\sin^*\frac{\phi_k}{2}\\
%     \cos^*\frac{\phi_k}{2}
%   \end{array}\right).
% \end{equation}
The model has three phases: the trivial PT-unbroken gapped phase for $w-v<-u$, the PT-broken gapless phase for $-u<w-v<u$, and the topologically nontrivial PT-unbroken gapped phase for $w-v>u$~\cite{Chang2019}. 
\begin{figure}
    \centering
	\includegraphics[width=1\columnwidth]{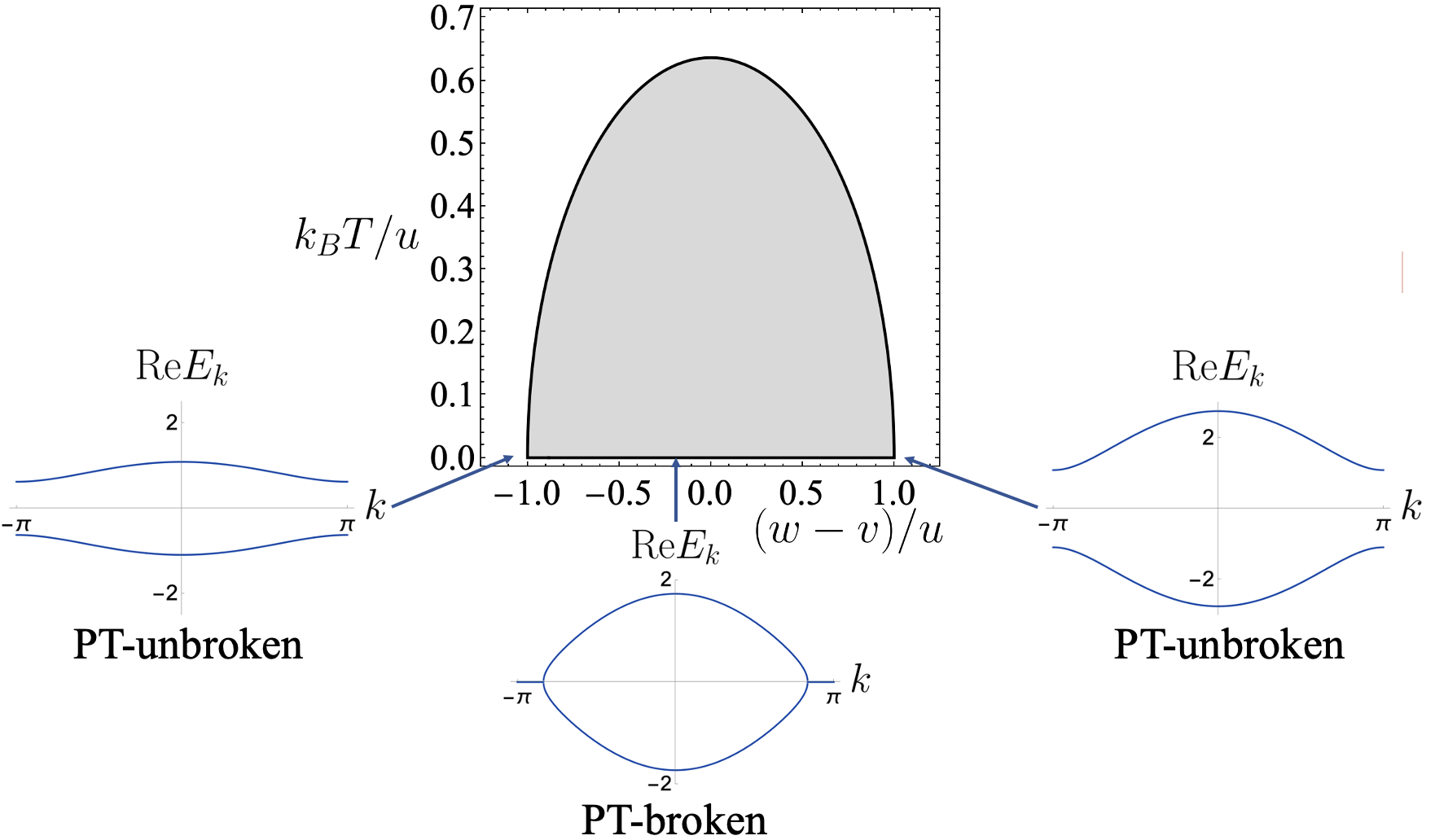}
	
	\caption{Distribution of Yang-Lee zeros (shaded region) of the SSH model on the plane of the temperature and the hopping parameters. At zero temperature, the onset of Yang-Lee zeros coincides with the quantum phase transitions indicated by the band structure (blue curves).}
	
	\label{SSH_zeros}
	\end{figure}
%On the phase boundary $|v-w|=u$, there exists the entanglement transition from the subarea law with $c=1$ in the gapless phase to the area law in the gapped phase. To understand the entanglement transition, we move to define the partition function to calculate the distribution of Yang-Lee zeros here. 
%We first introduce the linear transformation of the fermion operators as $\gamma_{kA}=\alpha_kc_{kA}+\beta_kc_{kB}$ and $\bar{\gamma}_{kA}=\bar{\alpha}_kc_{kA}+\beta_kc_{kB}$ (and similar equations also hold for $\gamma_{kB}$ and $\bar{\gamma}_{kB}$), where the coefficients satisfy $\alpha_k\bar{\alpha}_k+\beta_k^2=1$. (The explicit forms of the coefficients are shown in Supplemental Material~\cite{SupplementaryMaterial}). Under the linear transformation, the Hamiltonian \eqref{eq:second-quantization} can be diagonalized as
% \begin{equation}
%     H=\sum_kH_k=\sum_k(E_k\bar{\gamma}_{kA}\gamma_{kA}-E_k\bar{\gamma}_{kB}\gamma_{kB}).
% \end{equation}
% We note that since $v_k^*\neq\bar{v}_k$, $\bar{\gamma}_{kA(B)}=\gamma_{kA(B)}^{\dagger}$ does not hold in general. However, they still satisfy the fermionic anti-commutation relation: $\{\bar{\gamma}_{kA(B)},\gamma_{k'A(B)}\}=\delta_{k,k'}$. 

The partition function of the non-Hermitian SSH model is defined as~\footnote{Here we assume the chemical potential $\mu=0$.}
\begin{align}\label{eq:partition_function}
    Z&:=\sum_{n}\langle E_n^L|e^{-\beta H}|E_n^R\rangle.\nonumber\\
    &=\prod_k(1+e^{-\beta E_k})(1+e^{\beta E_k}),
\end{align}
where $\beta:=1/k_BT$ is the inverse temperature and the states $|E_n^{L(R)}\rangle$ are the left (right) eigenstates of the Hamiltonian with the energy $E_n$. The conditions of zeros of the partition function can be determined as
\begin{equation}\label{eq:zero-condition}
    \text{Re}[E_k]=0, \text{Im}[E_k]=(2n+1)\pi/\beta
\end{equation}
with $n\in\mathbb{Z}$. To figure out the distribution of Yang-Lee zeros, we first assume a finite $\beta$ and investigate where the condition \eqref{eq:zero-condition} is satisfied in the momentum space when we take the thermodynamic limit. Then we take the zero-temperature limit to see how the distribution of Yang-Lee zeros approaches the quantum phase transition points. The order of taking the two limits cannot be exchanged. In the zero-temperature limit $\beta\to\infty$, we find that the condition \eqref{eq:zero-condition} is satisfied in the thermodynamic limit for the gapless phase $|v-w|\leq u$, corresponding to the whole PT-broken phase. In the gapless phase, there always exists an interval in the momentum space with a finite purely imaginary part of the energy spectrum, which induces the Yang-Lee zeros. In Fig. \ref{SSH_zeros}, we show how Yang-Lee zeros approach the quantum phase transition points $|w-v|=u$ by taking the zero-temperature limit. To further understand the distribution of Yang-Lee zeros, we introduce the number of roots of the partition function $\chi$~\cite{Hongchao2023}, which is defined as the number of the integer $n$ satisfying the condition of zeros in Eq. \eqref{eq:zero-condition}. The number of roots $\chi$ can be understood as the density of Yang-Lee zeros on the plane of parameters. In the zero-temperature limit, the number of roots $\chi$ takes the form of
\begin{equation}
    \chi/\beta\simeq\frac{1}{2\pi}\sqrt{u^2-|v-w|^2},
\end{equation}
where $u>|v-w|$ in the PT-broken phase. Near the phase boundary $|v-w|=u$, the asymptotic behavior of $\chi$ can be shown as $\chi\propto(u-|v-w|)^{1/2}$ which is attributed to the exceptional points~\cite{Hongchao2023} in the gapless phase. Since the dispersion relation is $E_k^{\pm}=\pm\sqrt{v^2+w^2+2vw\cos k-u^2}$, the gap closes at 
 \begin{equation}
     k_E=\pm\arccos\frac{u^2-v^2-w^2}{2vw},
 \end{equation}
which correspond to the exceptional points. Around the exceptional points, the excitation spectrum takes the form of $E\propto(k-k_E)^{1/2}$, giving rise to the asymptotic behavior of $\chi$. 

Furthermore, we examine the Yang-Lee edge singularity by the correlation functions and its associated critical exponents. The correlation function
\begin{align}
    C_{\alpha\beta}(x)&={}_{L}\langle c_{\alpha}^{\dagger}(x)c_{\beta}(0)\rangle_{R},\nonumber\\
    &:=\frac{1}{Z}\langle G_L|c_{\alpha}^{\dagger}(x)c_{\beta}(0)e^{-\beta H}|G_R\rangle,
\end{align}
where $\alpha,\beta\in\{A,B\}$ indicating the sublattice index and the states $|G_{L(R)}\rangle$ represents the left (right) eigenstates of the Hamiltonian \eqref{eq:second-quantization} with only the negative real-energy modes filled. In the gapped phases $|v-w|>u$, this correlation function can be exactly calculated near the phase boundary at zero temperature as
\begin{equation}\label{eq:correlation}
    \lim_{x\to\infty}C_{\alpha\beta}(x)\propto\frac{e^{-x/\xi}}{\sqrt{x}},
\end{equation}
where $\delta=|v-w|-u$ and $\xi=vw/(2u\delta)$ is the correlation length. See Appendix \ref{sec:A} for details. Hence, from Eq. \eqref{eq:correlation}, we obtain the critical exponent $\nu=1$ which is defined as $\xi^{-1}\propto\delta^{\nu}$ \cite{Sachdev:2011uj}. In addition, from the correlation function \eqref{eq:correlation}, we have the anomalous scaling dimension $\eta=3/2$ defined as $\lim_{x\to\infty}C(x)\propto\exp(-x/\xi)/x^{d-2+\eta}$ \cite{Sachdev:2011uj}. The anomalous power $1/2$ originates from the dispersion relation $E\propto(k-k_E)^{1/2}$ near the exceptional points. Hence, combining with the dynamical critical exponent $z=1$, we have the critical exponents for the Yang-Lee universality class here, which are similar to the ones in the Yang-Lee universality class in the non-Hermitian BCS superconductivity \cite{Hongchao2023}. They are both induced by exceptional points in a PT-symmetric Hamiltonian. 

Now we turn to discuss the singularity induced by Yang-Lee zeros in terms of the entanglement entropy (EE) of the ground state. The EE of the subsystem $A$ is defined as $S_A:=\text{Tr}_{A}(\rho_A\ln\rho_A)$, where $\rho_A=\Tr_{\bar{A}}|G_R\rangle\langle G_L|$~\cite{Chang2019}. The EE discontinuously jumps from a subarea-law scaling $S_A=(1/6)\ln L_A$ in the PT-broken gapless phase, where Yang-Lee zeros are located, to an area-law scaling $S=\text{const}$ in the PT-unbroken gapped phase~\cite{Rudner2009,Chang2019}. Here $L_A$ denotes the length of the subsystem $A$. We can see that the edges of the distribution of zeros correspond to the entanglement transitions. We note that the correspondence is also applicable to the Hermitian SSH model. In the Hermitian case $u=0$, the system at the phase transition point $v=w$,  which is the edge of Yang-Lee zeros, exhibits a subarea-law scaling for the ground-state EE. To elucidate the generality of the correspondence, we consider an arbitrary local PT-symmetric free-fermion Hamiltonian. 
%In the momentum space, there are overall two regions: the PT-broken phase, where all eigenvalues are real, and the PT-unbroken phase, where all eigenvalues are imaginary and form complex conjugate pairs. The two regions are separated by the exceptional points. According to the conditions of Yang-Lee zeros in Eq. \eqref{eq:zero-condition}, we can see the Yang-Lee zeros appear in the PT-broken phase for general free fermions at absolute zero. 
By generalizing the discussion in Ref. \cite{Wolf2006}, we prove that the gapless PT-broken phase with Yang-Lee zeros shows at least a subarea-law scaling $S_A\geq c_{-}\ln L_A$ with $c_{-}>0$ and the gapped PT-unbroken phase shows an area-law scaling in the one-dimensional case. Therefore, the correspondence between the edges of the distribution of Yang-Lee zeros and the entanglement transitions is general for PT-symmetric free-fermion Hamiltonians. The Yang-Lee zeros introduce the singularity to the entanglement transition. Thus, entanglement entropy can be considered as a universal characterization of the Yang-Lee edge singularity in quantum phase transitions. See Appendix. \ref{sec: B} for the details of the proof.

\section{Interacting Model}
Above we have considered the Yang-Lee zeros in the quantum phase transitions of the non-interacting cases. In this part, we extend our discussion to an interacting case. We investigate the \emph{XXZ} model to understand the emergence of Yang-Lee zeros from the Bethe ansatz. The Hamiltonian of the \emph{XXZ} spin chain is defined as
\begin{equation}\label{eq:XXZ-Hamiltonian}
   H = - J \sum_{i = 1}^L (S_i^x S_{i + 1}^x + S_i^y S_{i + 1}^y + \Delta S_i^z S_{i + 1}^z),
\end{equation}
where $J>0$, $L$ is the size of the spin chain and we extend the anisotropy parameter $\Delta$ to a complex one to consider the distribution of Yang-Lee zeros. Here we focus on the ferromagnetic phase transition point $\Delta=1$. 
%At this phase transition point, the ground-state entanglement entropy experiences a change from the subarea law $S=\frac{1}{2}\log l$ \cite{Chen_2013} in the gapless phase $\Delta<1$ (near the phase boundary) to the area law in the gapped phase $\Delta>1$. 
In the Hermitian case, the ground state of the gapped (gapless) phase $\Delta>1$ ($-1<\Delta<1$) is (anti)ferromagnetic~\cite{takahashi2005thermodynamics}.

To consider the condition of Yang-Lee zeros, we follow the definition of the partition function in Eq. \eqref{eq:partition_function}: $Z=\sum_n{}_{L}\langle E_n|\exp(-\beta H)|E_n\rangle_R$. We show the following theorem:
\begin{theorem}\label{theorem1}
    The solutions $\Delta$ of $Z=0$ around the ferromagnetic phase transition point $\Delta=1$ are given by the algebraic equation
    \begin{equation}\label{eq:Yang-Lee-1}
        \sum_{M=0}^L z^{M(L-M)}=0
    \end{equation}
   in the zero-temperature limit $\beta\to\infty$ for an arbitrary length $L$ of the spin chain with $z:=\exp(\beta J(\Delta-1)/(L-1))$.
\end{theorem}
From Eq. \eqref{eq:Yang-Lee-1}, we deduce that overall, there are $N$ complex roots of $z$ which can be defined as $z_1,\cdots,z_N$, where $N=(L/2)^2$ for even $L$ and $N=(L^2-1)/4$ for odd $L$. By rewriting Eq. \eqref{eq:Yang-Lee-1} as $\sum_{M=1}^Na_Mz^M=0$, we have $a_0=2$ and $a_N=1$ for an even $L$ or $a_0=a_N=2$ for an odd $L$, indicating that both $z=0$ and $z=\infty$ are not the solutions of Eq. \eqref{eq:Yang-Lee-1}. Therefore, all the roots $z$ are finite that give rise to the zeros $\Delta_j$ on the complex plane of $\Delta$ as
\begin{equation}
    \Delta_j=1+\frac{L-1}{\beta J}\log(z_j)+i(L-1)\frac{2n\pi}{\beta J},
\end{equation}
where $n\in\mathbb{Z}$ and $j=1,\cdots,N$. We can see that the zeros form a line around the phase transition point. In the zero-temperature limit, those zeros are equivalent to $\Delta_j=1+i(L-1)\frac{2n\pi}{\beta J}$, which are simply the line $\text{Re}\Delta=1$. This indicates that the Yang-Lee zeros approach the real axis of $\Delta$ through the line $\text{Re}\Delta=1$. Below we outline the proof of Theorem \ref{theorem1}. 

\begin{figure}
	\includegraphics[width=1\columnwidth]{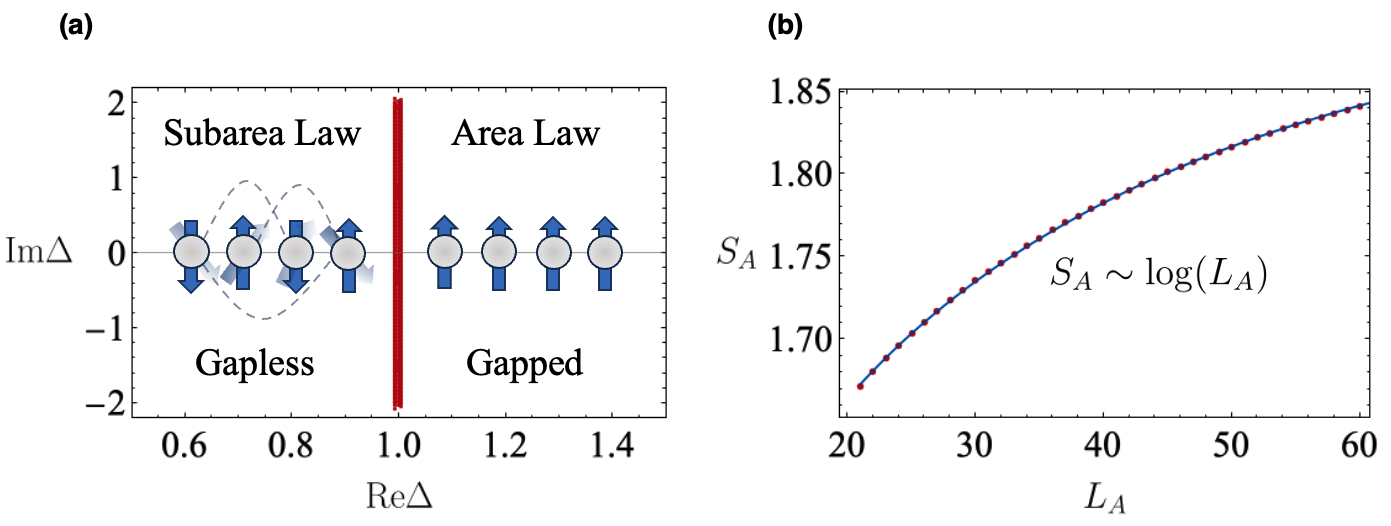}
	
	\caption{(a): The distribution line of Yang-Lee zeros in the non-Hermitian \emph{XXZ} model on the complex plane of $\Delta$. The size of the spin chain is $L=6$ and the inverse temperature is $\beta=100$. The zeros of the partition function approach the phase transition point $\Delta=1$ through a straight line. The dashed lines represent the entanglement between spins. Here we show the nearest 200 zeros to the phase transition point. (b): The scaling of EE $S_A$ of subsystem $A$ with size $L_A$ at the point $\Delta=0.99+0.01i$ where $L=200$. The fitting curve is chosen as $S_A=1.881+0.184\log(\sin(\pi L_A/L)))$, exhibiting the subarea law.}
	
	\label{XXZ_zero}
	\end{figure}

\begin{proof}
To begin with, we explain the structure of the exact solution of the \emph{XXZ} spin chain. The Bethe ansatz equations (BAEs) of the \emph{XXZ} chain are given by \cite{Yang1966,Yang_1966,takahashi2005thermodynamics}
\begin{equation}\label{eq:Bethe_XXZ}
  \left( \frac{\sin \frac{\phi}{2} (x_j + i)}{\sin \frac{\phi}{2} (x_j - i)}
  \right)^L = \prod_{l \neq j} \frac{\sin \frac{\phi}{2} (x_j - x_l + 2
  i)}{\sin \frac{\phi}{2} (x_j - x_l - 2 i)} ,
\end{equation}
where we define $x_j$ as $\exp (i k_j) = \sin \frac{\phi}{2} (x_j + i)/\sin \frac{\phi}{2}(x_j - i)$ and $\phi:=\cosh^{-1}\Delta$ with $k_j$ being the quasimomenta of the excited magnons. In Eq. \eqref{eq:Bethe_XXZ}, $j=1,\cdots,M$ where $M$ represents the number of magnons. The energy of the $M$-magnon sector takes the form of~\cite{takahashi2005thermodynamics}
\begin{align}\label{eq:energy_XXZ}
  E_M & =E_0 + \sum_{j = 1}^M J(\Delta-\cos k_j)\nonumber\\
  &= E_0 + \sum_{j = 1}^M \frac{J \sinh^2 \phi}{\cosh \phi - \cos \phi x_j},
\end{align}
where $E_0=-JL\Delta/4$ represents the ground state energy for the states with all spins up or down. Near the ferromagnetic phase transition point $\Delta=1$, we introduce a quantity $\delta:=\Delta-1$ and the quantity $\phi$ can be approximated as $\phi\approx\sqrt{2\delta}$. To understand the behavior of Yang-Lee zeros, we first turn to a limit $\delta\to0$, under which the system becomes the Heisenberg model. We can see the $(L+1)$-degeneracy of the ground state in the Heisenberg model. Those ground states can be constructed as $|M\rangle=(S^{-})^M|\Omega\rangle$ where we denote $|\Omega\rangle$ as the all-spin-up state. However, the $(L+1)$-degeneracy will be lifted if we add a perturbation to the parameter $\Delta$. To analyze the gap between levels, we consider the expansion of the BAEs for \emph{XXZ} chain \eqref{eq:Bethe_XXZ} around $\phi=0$,
\begin{equation}\label{eq:SBAE}
    L \cot \frac{\phi}{2} x_j = 2 \sum_{l \neq j} \cot \frac{\phi}{2} (x_j -
  x_l),
\end{equation}
which we call simplified Bethe ansatz equations (SBAEs) and are shown to be a set of algebraic equations of $\cot(\phi x_j/2)$. From the equations, one can prove that the solutions $\cot(\phi x_j/2)$ only depend on the size $L$. Meanwhile, the energy can also be expressed as a function of $\cot(\phi x_j/2)$,
\begin{equation}\label{eq:energy-3}
    E_M = E_0 + \sum_{j = 1}^M \frac{2 J \delta (1 + \cot^2 (\phi x_j))}{2 + \delta \cot^2 (\phi x_j/2)}.
\end{equation}
By substituting the solutions of the SBAEs into Eq. \eqref{eq:energy-3}, we obtain the energy of the $M$-magnon sector as
\begin{equation}\label{eq:energy-2}
        E_M=E_0+J\delta\frac{M(L-M)}{L-1}.
\end{equation}
With the definition $z=\exp(\beta J\delta/(L-1))$, we can rewrite the condition of $Z=0$ with Eq. \eqref{eq:energy-2} as
\begin{equation}\label{eq:zeros}
     \sum_{M=0}^L z^{M(L-M)}=0,
\end{equation}
 which is equivalent to Eq. \eqref{eq:Yang-Lee-1}. The maximal value of the exponent $M(L-M)$ is $L^2/4$ for an even $L$ with $M=L/2$ or $(L^2-1)/4$ for an odd $L$ with $M=(L-1)/2,(L+1)/2$. This completes the proof. Details of the proof are given in Appendix. \ref{sec3}
\end{proof}
To numerically verify the line distribution of the Yang-Lee zeros, we show the zeros of the partition function on the complex plane in Fig. \ref{XXZ_zero}(a) for $L=6$. 
We can further understand the line distribution of Yang-Lee zeros from the gap behavior of the non-Hermitian \emph{XXZ} model on the complex plane. For the Hamiltonians with $\text{Re}\Delta>1$, we can see that the systems must be gapped from Eq. \eqref{eq:energy-2}, with the gap $\Delta E=J\text{Re}\delta$. However, for those Hamiltonians with $\text{Re}\Delta<1$, the systems become gapless. In this phase, the ground state has $M=L/2$ magnons and the real part of the energy becomes $\text{Re}E=E_0+\frac{(L/2)^2}{L-1}J\text{Re}\delta$ with $\text{Re}\delta<0$ when $L$ is even. Then the energy level distance between the ground state and the first excited state is given by
\begin{equation}
    \Delta E=-\frac{J\text{Re}\delta}{L-1},
\end{equation}
which vanishes in the thermodynamic limit. Hence, the Hamiltonians with $\text{Re}\Delta<1$ behave as gapless systems. The behavior of the gap supports the existence of the Yang-Lee zeros near the phase transition point $\Delta=1$. 

Next, we examine the critical exponents for the Yang-Lee edge singularity of the non-Hermitian \emph{XXZ} model. We introduce a magnetic field $h$ to the Hamiltonian. By solving the BAE \eqref{eq:Bethe_XXZ}, the susceptibility in the gapless phase takes the form of $ \lim_{h\to0}\chi = \frac{2 \gamma}{J \pi \delta}$ near the critical point where $\gamma$ is a constant~\cite{takahashi2005thermodynamics}. Therefore, we obtain the critical exponent $\sigma=1$ from $\chi$ defined as $\chi\sim \delta^{-\sigma}$. At the critical point, the susceptibility diverges. In addition, by definition of the density of Yang-Lee zeros $g(\Delta)=\sum_{j}\delta(\Delta-\Delta_j)$, we have
 \begin{equation}
   g = \beta J \frac{N}{2 \pi (L - 1)} .
 \end{equation}
% \begin{equation}
%   \rho = \beta J \frac{(L / 2)^2}{2 \pi (L - 1)} .
% \end{equation}
% for even $L$ and 
% \begin{equation}
%   \rho = \frac{\beta J (L + 1)}{8 \pi} .
% \end{equation}
at the phase transition point. In the thermodynamic limit, the density has the scaling $g = \beta JL/8\pi$. 

Furthermore, here the edge of the Yang-Lee zeros is also associated with the entanglement transition of the ground state. The zeros are distributed on the line $\text{Re}\Delta=1$ and approach the phase transition point $\Delta=1$ when we decrease the temperature. Hence, we can consider the phase transition point as the edge of the Yang-Lee zeros. On the real axis, the EE of the gapless region $\text{Re}\Delta<1$ shows a subarea-law scaling while that of the gapped region $\text{Re}\Delta>1$ shows an area-law scaling. Therefore, the correspondence between the Yang-Lee edge singularity and the entanglement transitions is applicable to not only the non-interacting cases but also the interacting systems. 
%We note that the entanglement transition also happens on the complex plane since the structure of the real parts of the energy are the same as the ones on the real axis. Therefore, the gapless region on the complex plane should also show the subarea law. 
We note that the points on the distribution line of the Yang-Lee zeros also represent the entanglement transition points. To illustrate this, we calculate the EE of the model at $\Delta=0.99+0.01i$ with the non-Hermitian density-matrix renormalization group~\cite{Yamamoto2021,ITensor-r0.3} in Fig. \ref{XXZ_zero}(b), which also shows a subarea-law scaling. The reduced density matrix of the subsystem $A$ is defined as $\rho_A=\text{Tr}_{\bar{A}}|\psi_0^R\rangle\langle\psi_0^R|$, where $|\psi_0^R\rangle$ is the right eigenstate of the Hamiltonian \eqref{eq:XXZ-Hamiltonian} with the lowest real energy. On the other hand, the systems with $\text{Re}\Delta>1$ show the area law in the gapped phase since the ground state in this phase is a direct-product state. Therefore, all of the Yang-Lee zeros correspond to entanglement transition points on the complex plane. 

To generally understand how Yang-Lee zeros introduce the singularity in the entanglement transitions, we consider the density matrix $\rho=e^{-\beta H}/Z$ for the Hamiltonian $H(\lambda)$ with $\lambda$ being a parameter and assume the critical point of the system to be $\lambda=\lambda_c$ where we have the entanglement transition. We generalize the parameter $\lambda$ into a complex one and consider the Yang-Lee zeros on the complex plane of $\lambda$. Let the EE be $S_A=-\Tr_{A}[\rho_A\log\rho_A]$ where $\rho_A=\Tr_{\bar{A}}\rho$ is the reduced density matrix of the subsystem $A$ at absolute zero. 
% For those points where the system has $D$-degenerate ground states, we have $\rho_A=\sum_{\alpha=1}^D\Tr_{\bar{A}}[\exp(-i\beta\text{Im}[E_0^{\alpha}])|\psi_0^{\alpha R}\rangle\langle\psi_0^{\alpha L}|]/\tilde{Z}$ with 
% \begin{equation}
%     \tilde{Z}:=\exp(\beta\text{Re}[E_0])Z=\sum_{\alpha=1}^{D}\exp(-i\beta\text{Im}[E_0^{\alpha}]).
% \end{equation}
% The partition function $Z$ vanishes if the phase factors cancel with each other and hence Yang-Lee zeros emerge.
Since the zeros of the partition function also represent singularity of the EE, once one can find Yang-Lee zeros in the vicinity of $\lambda=\lambda_c$ on the complex plane of $\lambda$ at absolute zero, the singularity can enter the Hermitian system at $\lambda=\lambda_c$ and hence the entanglement transition takes place. The critical point $\lambda=\lambda_c$ is the edge of the distribution of Yang-Lee zeros.

\section{Conclusion and Outlook} In this work, we have investigated the distribution of Yang-Lee zeros in quantum phase transitions of the SSH model and the \emph{XXZ} model and shown how Yang-Lee zeros enter the phase transition points. In particular, in the \emph{XXZ} model, we have proposed the simplified Bethe ansatz equation (SBAE) to exactly solve the energy levels around the ground state and calculated the partition function in the zero-temperature limit. We have also explored the critical behavior at the edges of Yang-Lee zeros of both models. Furthermore, we find that the edges of Yang-Lee zeros at absolute zero correspond to the entanglement transition points of the ground state and discuss the mechanism behind the correspondence. We have also generalized the correspondence to an arbitrary PT-symmetric free-fermionic system.

The Yang-Lee zeros in quantum phase transitions studied in this paper are not only an interesting mathematical concept but also experimentally realizable. The non-Hermitian SSH model can be realized by dilating it into a larger Hermitian Hamiltonian due to the PT symmetry \cite{Zhang2021,Yu2022,Zeuner2015}. We can introduce an ancillary qubit as an environment coupled to the targeted spin to realize the Hamiltonian \eqref{eq:H_SSH}. For the non-Hermitian \emph{XXZ} model, we can realize it in a non-Hermitian Bose-Hubbard model in a strongly correlated regime \cite{Yamamoto2021,Zhihao2020,Tomita2017}, where the non-Hermiticity originates from the two-body loss in the open quantum systems. By coupling the target system to an ancilla, we expect the Yang-Lee zeros and its associated edge singularity to be measurable with the decoherence of the spin in the ancilla~\cite{Peng2014,Wei_2017,Wei_2018,Matsumoto2020}. The Yang-Lee zeros in those systems can also be measured from the diagonal elements of the time-evolution operator $U=\exp(i\tilde{H}t)$, where $\tilde{H}=iH$ and $t=\beta$~\cite{GaoPRL2024}. To be specific, we can measure the partition function of a non-Hermitian SSH model by preparing a state whose dynamics is described by the $2\times2$ Bloch Hamiltonian $iH_k$ in the momentum space, where $H_k$ is given in Eq. \eqref{eq:H_SSH}. Then the trace of the evolution operator gives the partition function: $Z=\prod_k\text{Tr}U_k$ with $t=\beta$ and $U_k=\exp{[i(iH_k)t]}=\exp{(-H_kt)}$. Similarly, we can also prepare a state whose dynamics is governed by the Hamiltonian $iH$ with tomography~\cite{Brien2004} to measure the partition function of the non-Hermitian \emph{XXZ} model where $H$ is shown in Eq. \eqref{eq:XXZ-Hamiltonian}. The partition function is given by $Z=\text{Tr}U=\text{Tr}\exp{[i(iH)t]}=\text{Tr}\exp{(-Ht)}$ where $t=\beta$. The transition of the entanglement entropy is expected to be observed in the quantum gas microscope~\cite{Islam2015}.

%While we have investigated the Yang-Lee zeros in the integrable systems, it is also worthwhile to explore how to express the entanglement entropy with the roots of the Bethe ansatz directly. We are also interested in whether the Yang-Lee zeros can be also observed in the other integrable fermionic models such as the one-dimensional Fermi-Hubbard model.

\section{ACKNOWLEDGMENTS}We are grateful to Zongping Gong, Masaya Nakagawa, Shinsei Ryu, Yunfeng Jiang, Kohei Kawabata, and Masahito Ueda for fruitful discussion. H. L. is supported by Forefront Physics and Mathematics Program to Drive Transformation (FoPM), a World-leading Innovative Graduate Study
(WINGS) Program, at the University of Tokyo. This work was also supported by JSPS KAKENHI Grants No. 24KJ0824.
%S.R. is supported by the National Science Foundation under Award No. DMR-2001181, and by a Simons Investigator Grant from the Simons Foundation (Award No. 566116).

\begin{widetext}
	\begin{appendix}
\section{Correlation Function in the SSH model}
\label{sec:A}

Here we consider the Yang-Lee zeros in the non-Hermitian SSH model. The Bloch Hamiltonian of non-Hermitian SSH model is given by \cite{Chang2019}
\begin{equation}
  H_k = i u \sigma_z + (v + w \cos k) \sigma_x + w \sin k \sigma_y =
  \left(\begin{array}{cc}
    i u & w e^{- i k} + v\\
    w e^{i k} + v & - i u
  \end{array}\right) .
\end{equation}
The energy spectrum is $E_{k, \pm} = \pm \sqrt{| w e^{- i k} + v |^2 - u^2}
\equiv \pm E_k$. Overall, there are two phases in the Hamiltonian: the
PT-symmetric phase for $| w - v | > u$ and the PT-broken phase for $| w - v | < u$.
In the PT-symmetric phase, the Hamiltonian is gapped with the energy gap
$\Delta = 2 \sqrt{(w - v)^2 - u^2}$. In the PT-broken phase, the Hamiltonian
is gapless with $\text{Re} [E_k] = 0$ between the exceptional points given by
\begin{equation}
  k_E = \pm \arccos \frac{u^2 - v^2 - w^2}{2 v w},
\end{equation}
where the gap closes and the Hamiltonian cannot be diagonalized. The
excitation spectrum around the exceptional points is given by $E_k \sim \sqrt{|
k - k_E |}$ \cite{Chang2019}.

Now we turn to calculate the critical exponents around the phase transition
points. We consider the two-point correlation function. To begin with, we
first diagonalize the Hamiltonian as
\begin{eqnarray}
  H & = & \sum_k\left(\begin{array}{cc}
    c_{k A}^{\dagger} & c^{\dagger}_{k B}
  \end{array}\right) \left(\begin{array}{cc}
    i u & v_k\\
    v_k^{\ast} & - i u
  \end{array}\right) \left(\begin{array}{c}
    c_{k A}\\
    c_{k B}
  \end{array}\right) \nonumber\\
  & = & \sum_k\left(\begin{array}{cc}
    \bar{\gamma}_{k A} & \bar{\gamma}_{k B}
  \end{array}\right) \left(\begin{array}{cc}
    \frac{v_k^{\ast}}{| v_k |} \cos \frac{\phi_k}{2} & \sin \frac{\phi_k}{2}\\
    - \frac{v_k^{\ast}}{| v_k |} \sin \frac{\phi_k}{2} & \cos \frac{\phi_k}{2}
  \end{array}\right) \left(\begin{array}{cc}
    i u & v_k\\
    v_k^{\ast} & - i u
  \end{array}\right) \left(\begin{array}{cc}
    \frac{v_k}{| v_k |} \cos \frac{\phi_k}{2} & - \frac{v_k}{| v_k |} \sin
    \frac{\phi_k}{2}\\
    \sin \frac{\phi_k}{2} & \cos \frac{\phi_k}{2}
  \end{array}\right) \left(\begin{array}{c}
    \gamma_{k A}\\
    \gamma_{k B}
  \end{array}\right) \nonumber\\
  & = & \sum_k\left(\begin{array}{cc}
    \bar{\gamma}_{k A} & \bar{\gamma}_{k B}
  \end{array}\right) \left(\begin{array}{cc}
    E_k & 0\\
    0 & - E_k
  \end{array}\right) \left(\begin{array}{c}
    \gamma_{k A}\\
    \gamma_{k B}
  \end{array}\right), 
\end{eqnarray}
where the transformation is given by
\begin{eqnarray}
  \gamma_{k A} & = & \frac{v_k^{\ast}}{| v_k |} \cos \frac{\phi_k}{2} c_{k A}
  + \sin \frac{\phi_k}{2} c_{k B}, \label{transform1} \\
  \gamma_{k B} & = & - \frac{v_k^{\ast}}{| v_k |} \sin \frac{\phi_k}{2} c_{k
  A} + \cos \frac{\phi_k}{2} c_{k B}, \\
  \bar{\gamma}_{k A} & = & \frac{v_k}{| v_k |} \cos \frac{\phi_k}{2} c_{k
  A}^{\dagger} + \sin \frac{\phi_k}{2} c^{\dagger}_{k B}, \\
  \bar{\gamma}_{k B} & = & - \frac{v_k}{| v_k |} \sin \frac{\phi_k}{2} c_{k
  A}^{\dagger} + \cos \frac{\phi_k}{2} c^{\dagger}_{k B} . \label{transform2} 
\end{eqnarray}
From the transformation, we have $\bar{\gamma}_{k A} \neq \gamma_{k A}^{\dagger}$
and $\bar{\gamma}_{k B} \neq \gamma_{k B}^{\dagger}$ since the Hamiltonian is non-Hermitian. However, the anti-commutation relations between $\gamma_{k A (B)}$ and $\bar{\gamma}_{k A (B)}$ still follow the fermionic ones, which can be proved as below:
\begin{eqnarray}
  \{ \bar{\gamma}_{k A}, \gamma_{k' A} \} & = & \left\{ \frac{v_k}{| v_k |}
  \cos \frac{\phi_k}{2} c_{k A}^{\dagger} + \sin \frac{\phi_k}{2}
  c^{\dagger}_{k B}, \frac{v_{k'}^{\ast}}{| v_{k'} |} \cos \frac{\phi_{k'}}{2}
  c_{k' A} + \sin \frac{\phi_{k'}}{2} c_{k' B} \right\} \nonumber\\
  & = & \cos^2 \frac{\phi_k}{2} \{ a_{k A}^{\dagger}, a_{k' A} \} + \sin^2
  \frac{\phi_k}{2} \{ a_{k B}^{\dagger}, a_{k' B} \} \nonumber\\
  & = & \delta_{k, k'}, \\
  \{ \bar{\gamma}_{k B}, \gamma_{k' B} \} & = & \left\{ - \frac{v_k}{| v_k |}
  \sin \frac{\phi_k}{2} c_{k A}^{\dagger} + \cos \frac{\phi_k}{2}
  c^{\dagger}_{k B}, - \frac{v_{k'}^{\ast}}{| v_{k'} |} \sin
  \frac{\phi_{k'}}{2} c_{k' A} + \cos \frac{\phi_{k'}}{2} c_{k' B} \right\}
  \nonumber\\
  & = & \sin^2 \frac{\phi_k}{2} \{ c_{k A}^{\dagger}, c_{k' A} \} + \cos^2
  \frac{\phi_k}{2} \{ c_{k B}^{\dagger}, c_{k' B} \} \nonumber\\
  & = & \delta_{k, k'}, \\
  \{ \bar{\gamma}_{k A}, \gamma_{k' B} \} & = & \left\{ \frac{v_k}{| v_k |}
  \cos \frac{\phi_k}{2} c_{k A}^{\dagger} + \sin \frac{\phi_k}{2}
  c^{\dagger}_{k B}, - \frac{v_k^{\ast}}{| v_k |} \sin \frac{\phi_k}{2} c_{k
  A} + \cos \frac{\phi_k}{2} c_{k B} \right\} \nonumber\\
  & = & - \cos \frac{\phi_k}{2} \sin \frac{\phi_k}{2} \{ c_{k A}^{\dagger},
  c_{k' A} \} + \cos \frac{\phi_k}{2} \sin \frac{\phi_k}{2} \{ c_{kB}^{\dagger}, c_{k' B} \} \nonumber\\
  & = & 0. 
\end{eqnarray}
Hence, we can still consider $\bar{\gamma}_{k A (B)}$ as the creation operator
of the fermion with the flavor $A (B)$ and the momentum $k$, similar to the non-Hermitian Bogoliubov transformation in Ref. \cite{Yamamoto2019}. Meanwhile, we can
calculate the correlation functions with the linear transformations
\eqref{transform1}-\eqref{transform2}. Here we define the correlation function as
\begin{equation}
  C_{\alpha \beta} (x) =_L \langle c_{\alpha}^{\dagger} (x) c_{\beta} (0)
  \rangle_R := \frac{1}{Z} \langle G_L | c_{\alpha}^{\dagger}
  (x) c_{\beta} (0) e^{- \beta E_{k s}} | G_R \rangle,
\end{equation}
where $\alpha, \beta \in \{ A, B \}$ and the state $|G_{L(R)}\rangle$ represent the right (left) eigenstate of the Hamiltonian with only the negative real-energy modes filled. It has four components which can be calculated from the Fourier transformation. We first show the forms of correlation functions in the momentum space as
\begin{eqnarray}
  C_{A A} (k) & = & _L \langle c_{k A}^{\dagger} c_{k A} \rangle_R \nonumber\\
  & = & _{L}\left\langle \left( \frac{v_k^{\ast}}{| v_k |} \cos \frac{\phi_k}{2}
  \bar{\gamma}_{k A} - \frac{v_k^{\ast}}{| v_k |} \sin \frac{\phi_k}{2}
  \bar{\gamma}_{k B} \right) \left( \frac{v_k}{| v_k |} \cos \frac{\phi_k}{2}
  \gamma_{k A} - \frac{v_k}{| v_k |} \sin \frac{\phi_k}{2} \gamma_{k B}
  \right) \right\rangle_R \nonumber\\
  & = & \cos^2 \frac{\phi_k}{2} {}_L\langle \bar{\gamma}_{k A} \gamma_{k A}
  \rangle_R + \sin^2 \frac{\phi_k}{2} {}_L\langle \bar{\gamma}_{k B} \gamma_{k B}
  \rangle_R \nonumber\\
  & = & \cos^2 \frac{\phi_k}{2} \frac{1}{e^{\beta E_k} + 1} + \sin^2
  \frac{\phi_k}{2} \frac{1}{e^{- \beta E_k} + 1},
  \end{eqnarray}
  \begin{eqnarray}
    C_{B B} (k) & = & _L \langle c_{k B}^{\dagger} c_{k B} \rangle_R \nonumber\\
  & = & {}_L\left\langle \left( \sin \frac{\phi_k}{2} \bar{\gamma}_{k A} + \cos
  \frac{\phi_k}{2} \bar{\gamma}_{k B} \right) \left( \sin \frac{\phi_k}{2}
  \gamma_{k A} + \cos \frac{\phi_k}{2} \gamma_{k B} \right) \right\rangle_R
  \nonumber\\
  & = & \sin^2 \frac{\phi_k}{2} {}_L\langle \bar{\gamma}_{k A} \gamma_{k A}
  \rangle_R + \cos^2 \frac{\phi_k}{2} {}_L\langle \bar{\gamma}_{k B} \gamma_{k B}
  \rangle_R \nonumber\\
  & = & \sin^2 \frac{\phi_k}{2} \frac{1}{e^{\beta E_k} + 1} + \cos^2
  \frac{\phi_k}{2} \frac{1}{e^{- \beta E_k} + 1}, \\
    C_{A B} (k) & = & _L \langle c_{k A}^{\dagger} c_{k B} \rangle_R \nonumber\\
  & = & {}_L\left\langle \left( \frac{v_k^{\ast}}{| v_k |} \cos \frac{\phi_k}{2}
  \bar{\gamma}_{k A} - \frac{v_k^{\ast}}{| v_k |} \sin \frac{\phi_k}{2}
  \bar{\gamma}_{k B} \right) \left( \sin \frac{\phi_k}{2} \gamma_{k A} + \cos
  \frac{\phi_k}{2} \gamma_{k B} \right) \right\rangle_R \nonumber\\
  & = & \frac{v_k^{\ast}}{| v_k |} \cos \frac{\phi_k}{2} \sin
  \frac{\phi_k}{2} {}_L\langle \bar{\gamma}_{k A} \gamma_{k A} \rangle_R -
  \frac{v_k^{\ast}}{| v_k |} \sin \frac{\phi_k}{2} \cos \frac{\phi_k}{2}
  {}_L\langle \bar{\gamma}_{k B} \gamma_{k B} \rangle_R \nonumber\\
  & = & \frac{v_k^{\ast}}{| v_k |} \cos \frac{\phi_k}{2} \sin
  \frac{\phi_k}{2} \left( \frac{1}{e^{\beta E_k} + 1} - \frac{1}{e^{- \beta
  E_k} + 1} \right), 
\end{eqnarray}
\begin{eqnarray}
  C_{B A} (k) & = & _L \langle c_{k B}^{\dagger} c_{k A} \rangle_R \nonumber\\
  & = & {}_L\left\langle \left( \sin \frac{\phi_k}{2} \bar{\gamma}_{k A} + \cos
  \frac{\phi_k}{2} \bar{\gamma}_{k B} \right) \left( \frac{v_k}{| v_k |} \cos
  \frac{\phi_k}{2} \gamma_{k A} - \frac{v_k}{| v_k |} \sin \frac{\phi_k}{2}
  \gamma_{k B} \right) \right\rangle_R \nonumber\\
  & = & \frac{v_k}{| v_k |} \cos \frac{\phi_k}{2} \sin \frac{\phi_k}{2}{}_L\langle \bar{\gamma}_{k A} \gamma_{k A} \rangle_R - \frac{v_k}{| v_k |} \sin\frac{\phi_k}{2} \cos \frac{\phi_k}{2} {}_L\langle \bar{\gamma}_{k B} \gamma_{kB} \rangle_R \nonumber\\
  & = & \frac{v_k}{| v_k |} \cos \frac{\phi_k}{2} \sin \frac{\phi_k}{2}
  \left( \frac{1}{e^{\beta E_k} + 1} - \frac{1}{e^{- \beta E_k} + 1} \right), 
\end{eqnarray}

Here we only consider the $\text{LR}$ correlation function and set the chemical potential $\mu=0$. Then we consider the
correlation function in the real space by Fourier transformation. In the gapped phase we have
\begin{eqnarray}
  C_{A A} (x) & = & \int \frac{d k}{2 \pi} \langle c_{k A}^{\dagger} c_{k A}
  \rangle e^{i k x} \nonumber\\
  & = & \int \frac{d k}{2 \pi} \left( \cos^2 \frac{\phi_k}{2}
  \frac{1}{e^{\beta E_k} + 1} + \sin^2 \frac{\phi_k}{2} \frac{1}{e^{- \beta
  E_k} + 1} \right) e^{i k x} \nonumber\\
  & \xrightarrow{T = 0 } & \int \frac{d k}{2 \pi} \frac{1}{2} \left( 1 -
  \frac{i u}{\sqrt{| v_k |^2 - u^2}} \right) e^{i k x} \nonumber\\
  & = & - \frac{i u}{4 \pi} \int_{- \pi}^{\pi} \frac{1}{\sqrt{(v - w)^2 - u^2
  + 2 v w (1 + \cos k)}} e^{i k x} d k. 
\end{eqnarray}
This integral can approximately be calculated near the critical point. Near the critical points $| v - w | = u$, we have
\begin{equation}
  C_{A A} (x) = - \frac{i u}{4 \pi} \int_{- \pi}^{\pi} \frac{1}{\sqrt{2 u
  \delta + 2 v w (1 + \cos k)}} e^{i k x} d k.
\end{equation}
In this integral, we find that the parts around the points $k = \pm \pi$ dominate. Hence, we expand around the two points and obtain~\footnote{Here we
only take the critical point $v - w = u$ as an example.}

\begin{eqnarray}
  C_{A A} (x) & \simeq & - e^{i \pi x} \frac{i u}{2 \pi} \int_0^{\infty}
  \frac{\cos (k x)}{\sqrt{2 u \delta + v w k^2}} d k. \nonumber\\
  & = & - e^{i \pi x} \frac{i u}{2 \pi \sqrt{2 u \delta}} \int_{-
  \infty}^{\infty} \frac{\cos (k x)}{\sqrt{1 + \xi^2 k^2}} d k \nonumber\\
  & = & - e^{i \pi x} \frac{i u}{2 \pi \sqrt{2 u \delta} \xi} K_0 \left(
  \frac{x}{\xi} \right), 
\end{eqnarray}
where $\delta = v - w - u$ and $\xi = vw / 2 u \delta$. Here $K_0 (x)$ denotes the modified Bessel function of the second kind. Similarly, we also have
\begin{eqnarray}
  C_{B B} (x) & \simeq & e^{i \pi x} \frac{i u}{2 \pi} \int_0^{\infty}
  \frac{\cos (k x)}{\sqrt{2 u \delta + 4 v w k^2}} d k. \nonumber\\
  & = & e^{i \pi x} \frac{i u}{2 \pi \sqrt{2 u \delta}} \int_{-
  \infty}^{\infty} \frac{\cos (k x)}{\sqrt{1 + \xi^2 k^2}} d k \nonumber\\
  & = & e^{i \pi x} \frac{i u}{2 \pi \sqrt{2 u \delta} \xi} K_0 \left(
  \frac{x}{\xi} \right) . 
\end{eqnarray}
\begin{eqnarray}
  C_{A B} (x) & = & \int \frac{d k}{2 \pi} \left[ \frac{v_k^{\ast}}{| v_k |}
  \cos \frac{\phi_k}{2} \sin \frac{\phi_k}{2} \left( \frac{1}{e^{\beta E_k} +
  1} - \frac{1}{e^{- \beta E_k} + 1} \right) \right] e^{i k x} \nonumber\\
  & \xrightarrow{T = 0} & - \frac{1}{2} \int \frac{d k}{2 \pi} \sin \phi_k
  e^{i k x} \nonumber\\
  & = & - \frac{1}{2} \int \frac{d k}{2 \pi} \sqrt{1 + \frac{u^2}{v^2 + w^2 +
  2 v w \cos k - u^2}} e^{i k x} \nonumber\\
  & \simeq & - \int_{- \infty}^{\infty} \frac{d k}{2 \pi} \sqrt{1 + \frac{u /
  2 \delta}{1 + \xi^2 k^2}} \cos (k x) \nonumber\\
  & \simeq & - \frac{1}{2 \pi \xi} \sqrt{\frac{2 \delta}{u}} K_0 \left(
  \frac{x}{\xi} \right) . 
\end{eqnarray}
The correlation function $C_{B A} (x)$ shows the same result as $C_{A B} (x)$. Therefore, all the correlation functions have the same long-range scaling behavior with the distance $x$ as
\begin{equation}
  \lim_{x \rightarrow \infty} C_{\alpha \beta} (x) \sim \frac{e^{- x /
  \xi}}{\sqrt{x}},
\end{equation}
which originates from the scaling behavior of $K_0 (x)$. This exponential decay is attributed to the finite energy gap of the system. From the definition of the critical exponents in the main text, we obtain $\nu=1,\eta=3/2$.

\section{Entanglement Entropy in Non-interacting Fermionic Systems}
\label{sec: B}

In this section, we calculate the entanglement entropy for the general local non-interacting fermionic systems in the $d$-dimensional case with PT symmetry. Here we consider the entanglement entropy of the ground-state density matrix $\rho=|\psi_0^R\rangle\langle\psi_0^L|$, where $|\psi_0^{R(L)}\rangle$ represents the right (left) ground state of the system. The entanglement entropy of the subsystem $A$ is defined as
\begin{equation}
    S_A=-\text{Tr}[\rho_A\log\rho_A],\rho_A:=\text{Tr}_{\bar{A}}\rho.
\end{equation}
The ground state here is defined as the eigenstate with only the negative-energy modes filled. Here we consider the case of ground-state degeneracy where the system has two phases: the PT-broken phase and the PT-unbroken phase. The system has pure imaginary eigenvalues in the PT-broken phase which indicates that the phase is gapless and only finite real values in the PT-unbroken phase which indicates that the phase is gapped. In the PT-broken phase, the pure imaginary eigenvalues are surrounded by the exceptional points in the Brillouin zone, namely exceptional rings~\cite{Budich2019,Bergholtz2021}. The partition function defined in Eq. (5) in the main text can be written as
\begin{equation}
    Z=\prod_{k,\alpha}\left(1+e^{-\beta E_k^{\alpha}}\right),
\end{equation}
where $\alpha$ is the index of band. In the PT-broken phase, since $E_k^{\alpha}$ can be imaginary values for some given $\alpha$, we have $Z=0$ for those momenta satisfying
\begin{equation}\label{eq: condition}
    \text{Im}E_k^{\alpha}=\frac{(2n+1)\pi}{\beta}.
\end{equation}
We can infer that there always exist momenta $k$s satisfying Eq. \eqref{eq: condition} at absolute zero in the PT-unbroken phase, where Yang-Lee zeros emerge. Next, we are going to prove that the entanglement entropy shows at least a subarea-law scaling in the PT-unbroken phase. To show this, we follow the method used in Ref. \cite{Wolf2006}. The correlation matrix is defined as
\begin{equation}
    \gamma_{i\alpha,j\beta}=\delta_{ij}\delta_{\alpha\beta}-2\text{Tr}[\rho c^{\dagger}_{i\alpha}c_{j\beta}]=\langle\psi_0^L|(I_{ij}I_{\alpha\beta}-2c^{\dagger}_{i\alpha}c_{j\beta})|\psi_0^R\rangle,
\end{equation}
where $i,j\in\mathbb{Z}$ are the positions of the fermions in the subsystem $A$, $\alpha,\beta$ are the band indices and $I$ is the identity matrix. To simplify the calculation, here we assume that the subsystem is a hypercube with edge length $L$. The entanglement entropy can be expressed with the eigenvalues $\lambda_M$ of the correlation matrix in the subsystem $A$~\cite{Chang2019}.
\begin{align}
    S_A&=\sum_{M=1}^{NL^d}h(\lambda_{M}),\\
    h(x)&=-\frac{1+x}{2}\log\left(\frac{1+x}{2}\right)-\frac{1-x}{2}\log\left(\frac{1-x}{2}\right),
\end{align}
where $N$ is the number of the energy bands. We begin from a two-band case. We consider the case where the Fermi surface is the exceptional ring and the correlation matrix is given by
\begin{equation}
    \gamma_{i\alpha,j\beta}=\int d^dk(1-2\eta(k))e^{ik(i-j)}\delta_{\alpha\beta},
\end{equation}
where $\eta(k):=\theta(-\text{Re}E_k+\epsilon)$ is an indicator function with $\epsilon>0$ an infinitesimal number. We note that here we choose the right and left ground states as $|\psi_0^R\rangle=\prod_{k:\text{Re}[E_k^{\alpha}]< 0}c_{k\alpha}^{\dagger}|\mathrm{vac}\rangle$ and $|\psi_0^L\rangle=\prod_{k:\text{Re}[E_k^{\alpha}]< 0}\langle \mathrm{vac}|c_{k\alpha}$. By utilizing the inequality
\begin{equation}
    h(x)>\frac{1}{2}(1-x^2),
\end{equation}
we have the lower bound of the entanglement entropy
\begin{equation}
    S_A>\frac{1}{2}\text{Tr}[I-\gamma^2].
\end{equation}
With the introduction of the kernel~\cite{Wolf2006}
\begin{equation}
    F_L(x)=\sum_{i,j\in\mathbb{Z}_L}e^{ix(i-j)}=\frac{\cos(Lx)-1}{\cos(x)-1},
\end{equation}
we have
\begin{align}
    \text{Tr}[I-\gamma^2]&=\frac{4}{(2\pi)^{2d}}\int d^dk\int d^dk'\eta(k)(1-\eta(k'))F_L(k-k')\nonumber\\
    &=\frac{4}{(2\pi)^{2d}}\int d^dq \Xi(q)F_L(q),
\end{align}
where
\begin{equation}
    \Xi(q):=\int d^dk\eta(k)[1-\eta(q+k)].
\end{equation}
Similarly, we can bound the function $\Xi(q)$ with $s_{-}\|q\|_2$ where
\begin{equation}
    s_{-}=\frac{1}{2}A_{\text{FS}}
\end{equation}
with $A_{\text{FS}}$ defined as the area of the Fermi surface since the condition for $\text{Re}E_k=0$ and that for $\text{Im}E_k=0$ are the same here. Therefore, we obtain
\begin{align}
    \int d^dq s_{-}\|q\|_2F_L(q)&\geq\frac{1}{\sqrt{d}}\int d^dq s_{-}\|q\|_1F_L(q)\nonumber\\
    &=\sum_{i=1}^{d}\frac{1}{\sqrt{d}}\int d^dq s_{-}F_L(q)q_i\nonumber\\
    &=\sqrt{d}s_{-}(2\pi L)^{d-1}\int_{0}^{\pi} dxF_L(x)x.
\end{align}
In the thermodynamic limit, the integral $\int_{0}^{\pi} dxF_L(x)x$ scales as $2\log L$. Hence, the entanglement entropy is lower bounded by
\begin{equation}
    S_A>\frac{2\sqrt{d}}{(2\pi)^{d+1}}s_{-}L^{d-1}\log L:=c_{-}L^{d-1}\log L,
\end{equation}
where $c_{-}:=\frac{2\sqrt{d}}{(2\pi)^{d+1}}s_{-}$. The entanglement entropy shows at least a logarithmic scaling when $d=1$. For the multi-band cases, we have $S_A>\lfloor N/2\rfloor c_{-}\log L$, where $\lfloor x\rfloor$ is the floor function. Hence, the PT-unbroken phase with Yang-Lee zeros is always accompanied by at least a subarea-law scaling for the entanglement entropy. In the PT-broken phase, the entanglement entropy exhibits area law since it is gapped. Hence, the edges of the distribution of Yang-Lee zeros in one-dimensional cases always correspond to the entanglement transition. For the other cases where the Fermi surface is not the exceptional ring, the discussion is exactly the same as the Hermitian one~\cite{Wolf2006}. Since Yang-Lee zeros can only take place in the gapless PT-broken phase, the edge of the Yang-Lee zeros can always correspond to the entanglement transitions.

\section{Derivation of Eqs. \eqref{eq:energy-2} and \eqref{eq:zeros}}\label{sec3}

We here consider the one-dimensional (1D) \emph{XXZ} model with the Hamiltonian
\begin{equation}
  H = - J \sum_{i = 1}^L (S_i^x S_{i + 1}^x + S_i^y S_{i + 1}^y + \Delta S_i^z
  S_{i + 1}^z), \label{XXZ}
\end{equation}
where $J > 0$ and $\Delta$ is the anisotropy parameter. This model can be exactly solved by the Bethe ansatz. The Bethe ansatz equations of the \emph{XXZ} model are given by \cite{takahashi2005}
\begin{equation}
  \left( \frac{\sin \frac{\phi}{2} (x_j + i)}{\sin \frac{\phi}{2} (x_j - i)}
  \right)^L = \prod_{l \neq j} \frac{\sin \frac{\phi}{2} (x_j - x_l + 2
  i)}{\sin \frac{\phi}{2} (x_j - x_l - 2 i)} . \label{Bethe-XXZ}
\end{equation}
Here we define $\cosh \phi : = \Delta$ and $\exp (i k_j) = \sin \frac{\phi}{2}
(x_j + i) / \sin \frac{\phi}{2} (x_j - i)$, where $k_j$ is the quasimomentum. The index is given by $j = 1, \cdots, M$ with
$M$ being the number of the excited magnons. If we take the limit $\delta:=\Delta-1\rightarrow 0$, we can approximate $\phi$ as $\phi \simeq \sqrt{2 \delta}$.
When $\phi = 0$ ($\Delta$ = 1), Eq. \eqref{Bethe-XXZ} can be reduced as
\begin{equation}
  \left( \frac{x_j + i}{x_j - i} \right)^L = \prod_{l \neq j} \left( \frac{x_j
  - x_l + 2 i}{x_j - x_l - 2 i} \right), \label{Bethe-XXX}
\end{equation}
which consists of the Bethe ansatz equations of the Heisenberg model. In this case the variable
$x_j$ can be rewritten as $x_j = \cot (k_j / 2)$. From Eq. \eqref{Bethe-XXX}, we can see that $x_j = \infty
(k_j = 0)$ is a set of solutions of the Bethe ansatz, which represents that all
magnons have zero momentum or equal-weight superposition on each site of the
system. These states are ground states of the Heisenberg model deduced from
the energy:
\begin{equation}
  E_M = E_0 + \sum_{j = 1}^M J (1 - \cos k_j) = E_0 + \sum_{j = 1}^M \frac{2
  J}{x_j^2 + 1} .
\end{equation}

For the Heisenberg model with a system size $L$, there is $(L + 1)$-degeneracy
of the ground state, originating from the SU(2) symmetry of the model. These
eigenstates can be expressed as $| M \rangle = (S^-)^M | \Omega \rangle$ where
$S^- = \sum_j S_j^-$. However, when we have a finite $\phi$ for anisotropy, the degeneracy will be lifted. To be specific, we begin from the expansion of the Bethe ansatz equations \eqref{Bethe-XXZ} of the \emph{XXZ} model. Since we only consider the
distribution of Yang-Lee zeros around the phase transition point $\Delta = 1$,
we take the limit $\phi \rightarrow 0$ and expand the equation as
\begin{equation}
  \left( \frac{\sin \frac{\phi}{2} x_j + i \frac{\phi}{2} \cos \frac{\phi}{2}
  x_j}{\sin \frac{\phi}{2} x_j - i \frac{\phi}{2} \cos \frac{\phi}{2} x_j}
  \right)^L = \prod_{l \neq j} \frac{\sin \frac{\phi}{2} (x_j - x_l) + i \phi
  \cos \frac{\phi}{2} (x_j - x_l)}{\sin \frac{\phi}{2} (x_j - x_l) - i \phi
  \cos \frac{\phi}{2} (x_j - x_l)} .
\end{equation}
Up to the first order of $\phi,$ this expansion can be rewritten as
\begin{equation}
  L \cot \frac{\phi}{2} x_j = 2 \sum_{l \neq j} \cot \frac{\phi}{2} (x_j -
  x_l), \label{eq:Bethe}
\end{equation}
which is referred to as simplified Bethe ansatz equations (SBAEs). If we define $y_j =
\frac{\phi}{2} x_j$, the SBAEs can be rewritten as
\begin{equation}
  L \cot y_j = 2 \sum_{l \neq j} \cot (y_j - y_l) .
\end{equation}
We can see that this equation is purely an algebraic equation on $\cot y_j$, where the solutions only depend on the size $L$ and the magnon number $M$.
Meanwhile, the energy of the $M$-magnon excitation is given by
\begin{eqnarray}
  E = E_0 + \sum_{j = 1}^M \frac{J \sinh^2 \phi}{\cosh \phi - \cos \phi x_j} &
  = & E_0 + \sum_{j = 1}^M \frac{2 J \delta}{1 + \delta - \cos \phi x_j}
  \nonumber\\
  & = & E_0 + \sum_{j = 1}^M \frac{2 J \delta (1 + \cot^2 y_j)}{2 + \delta
  \cot^2 y_j} . \label{eq:energy} 
\end{eqnarray}
In the expansion above, we first assume that $\cot y_j$ is finite so that we can
expand the Bethe ansatz equation \eqref{Bethe-XXZ} up to the first order of
$\phi$. Now we can also see this point from the expression \eqref{eq:energy}.
Divergence of the factor $\cot y_j$ will increase $2 J$ to the energy,
indicating that this solution does not originate from the original $(L +
1)$-degeneracy of the Heisenberg spin chain. Therefore, we can exactly obtain all the deviated ground states from the SBAE. The finiteness of $\cot y_j$ of the deviated ground states can also be seen from the SBAE \eqref{eq:Bethe}. We
assume that there are divergences of $\cot y_j$ first. Then there must be a
series of $\cot y_j$ which satisfy $\cot y_j = a_j N$, where $j = 1, \cdots, m$, with $N$ being a divergent number and $a_j$ being a constant. Since some of the solutions $\cot y_j$ may not diverge, we always have $m \leqslant M
\leqslant L$. Then the SBAE becomes
\begin{equation}
  L a_j = 2 \sum_{l \neq j} \frac{a_j a_l}{a_l - a_j} .
\end{equation}
By summing up all the equalities, we have
\begin{equation}
  m L = 2 \frac{m (m - 1)}{2} \Rightarrow L = m - 1.
\end{equation}
However, this contradicts $m \leqslant L$. Therefore, all of the $\cot y_j$ should be convergent.

We then move to prove Eq. (16) in the main text. For convenience, we define a
new variable $\zeta_j = \cot y_j$. Then the SBAE can be rewritten as
\begin{equation}
  L \zeta_j = 2 \sum_{l \neq j} \frac{1 + \zeta_l \zeta_j}{\zeta_l - \zeta_j}
  . \label{eq:SBAE2}
\end{equation}
Then we have
\begin{eqnarray}
  \sum_j \zeta_j^2 & = & \frac{2}{L} \sum_{j, l \neq j} \frac{1 + \zeta_l
  \zeta_j}{\zeta_l - \zeta_j} \zeta_j \nonumber\\
  & = & - \frac{2}{L} \sum_{j, l \neq j} \frac{1 + \zeta_l \zeta_j}{\zeta_l -
  \zeta_j} \zeta_l, 
\end{eqnarray}
where we exchange the label $l, j$ in the second equality. Hence, by summation
of the right-hand side of the two equalities, we obtain
\begin{eqnarray}
  \sum_j \zeta_j^2 & = & - \frac{1}{L} \sum_{j, l \neq j} (1 + \zeta_l
  \zeta_j) \nonumber\\
  & = & - \frac{M (M - 1)}{L} - \frac{1}{L} \sum_{j, l \neq j} \zeta_l
  \zeta_j \nonumber\\
  & = & - \frac{M (M - 1)}{L} - \frac{1}{L} \left( \sum_j \zeta_j \right)^2 +
  \frac{1}{L} \sum_j \zeta_j^2 \nonumber\\
  & = & - \frac{M (M - 1)}{L} + \frac{1}{L} \sum_j \zeta_j^2, 
\end{eqnarray}
where we use $\sum_j \zeta_j = 0$ from Eq. \eqref{eq:SBAE2}. Hence, the
summation of $\zeta_j^2$ can be simplified as
\[ \sum_j \zeta_j^2 = - \frac{M (M - 1)}{L - 1} . \]
By revisiting the expression of energy \eqref{eq:energy}, we have
\begin{equation}
  E = E_0 + \sum_{j = 1}^M J \delta (1 + \cot^2 y_j) = E_0 + J \delta \left( M
  + \sum_{j = 1}^M \zeta_j^2 \right) = E_0 + J \delta \frac{M (L - M)}{L - 1}
  , \label{eq:energy2}
\end{equation}
which can indicate that the energy reaches the
maximum for $M = \frac{L}{2} \left( \frac{L - 1}{2} \right)$ when $L$ is even (odd). With the expression of energy \eqref{eq:energy2}, we show the
expression of the partition function at absolute zero as
\begin{equation}
  Z = \lim_{T \rightarrow 0} \text{Tr} e^{- \beta H} = \lim_{T \rightarrow 0}
  \sum_{M = 0}^L e^{- \beta E_M} = \lim_{T \rightarrow 0} e^{- \beta E_0}
  \sum_{M = 0}^L z^{M (L - M)} .
\end{equation}
Hence, the condition of $Z = 0$ is equivalent to
\begin{equation}
  \sum_{M = 0}^L z^{M (L - M)} = 0,
\end{equation}
which is exactly Eq. (17) or Eq. (10) in the main text. By solving this equation, we can obtain the distribution of Yang-Lee zeros, which is given by $\text{Re} \Delta = 1$ on the complex plane of $\Delta$.

Finally, we discuss the Yang-Lee edge singularity associated with the Yang-Lee zeros. Around the phase transition point $\Delta = 1$, we consider the susceptibility~\cite{takahashi2005}
\begin{equation}
  \chi = \frac{2 s_z}{h} = \frac{4 \gamma}{J \pi (\pi - \gamma) \sin \gamma}
  [1 + O (h^2) + O (h^{4 \gamma / (\pi - \gamma)})],
\end{equation}
where
\begin{equation}
  \cos \gamma = - \Delta .
\end{equation}
Around the ferromagnetic phase transition point, we have $\gamma = \pi$, which
shows the divergence of the susceptibility. The scaling of $\gamma$ near
$\Delta = 1$ is $\gamma \approx \pi - i \sqrt{2 \delta}$. Then we generalize
$\delta$ into an imaginary number where Yang-Lee zeros are located. The scaling of
the susceptibility is given by
\begin{equation}
  \chi (h = 0) \simeq \frac{4 \gamma}{J \pi \left( - i \sqrt{2 \delta} \right)
  i \sqrt{2 \delta}} = \frac{2}{J \delta} \propto \delta^{- 1} .
\end{equation}
Furthermore, we calculate the correlation length. Due to the finite deviation
$\delta$, the ground-state degeneracy is lifted and the system becomes gapped. Near the ferromagnetic phase transition point, the gap $\Delta E$ is given by
\begin{equation}
  \Delta E = \text{Re} [J \delta],
\end{equation}
which is determined by the Bethe ansatz. Hence, from the solution we can see the critical exponent $\varphi = 1$, where we define $\Delta E = J \times \text{Re} [\delta]^{\varphi}$.

    \end{appendix}
\end{widetext}

\bibliography{MyCollection.bib}

\end{document}